# Electronic transport under hydrostatic pressure and peculiar properties of the impurity energy spectrum in semiconductor arsenides of n-type: InAs, CdSnAs$_2$, CdGeAs$_2$ and GaAs

R.R. Bashirov*, S.F. Gabibov, M.I. Daunov, A.Yu. MollaeV

*Amirkhanov Institute of Physics, Dagestan Scientific Center of RAS, Makhachkala, 367003, Russia*

E-mail: rustam.bashirov@wayne.edu

**Abstract**

The energy spectrum of vacancies in n-type bulk crystals of undoped arsenides: InAs, GaAs, CdSnAs$_2$ and CdGeAs$_2$ has been investigated upon the data on pressure and temperature dependences of kinetic coefficients. It is concluded, that deep donor levels correspond to native vacancies of arsenic in these semiconductor materials. The vacancy level positions in the energy scale relatively to the conductivity band edge and their pressure coefficients are defined.

## 1. Introduction

It is known [1,2], that the electronic type of conductivity in undoped binary III-V and ternary II-IV-V$_2$ semiconductor arsenides InAs, GaAs, CdSnAs$_2$, and CdGeAs$_2$ is caused by a structure nonstoichiometry - vacancies in anionic sublattice, as arsenic is a highly volatile component. This conclusion has been confirmed by researches of the nature of defects in arsenides crystals irradiated with electrons, where deep donor centers were found [3]. Unlike the energy of shallow impurity centers that "follow" the band they are genetically connected with, the energy of the deep impurity centers concerning the absolute vacuum level under isotropic compression of a crystal lattice, remains practically a constant [4-7]. This invariance is due to the fact, that wave functions of localized states should be built in the entire Brillouin zone, and the influence of all-round pressure on energy of states is defined by evolution of total appearance of the energy spectrum, and not only by the nearest one or two bands [8-10].

It is pertinently to notice, that the identification of an impurity center to be shallow or deep merely from data about energy of ionization, capture section, and other phenomenological characteristics is problematically. Complementary researches on modification of the charge carrier energy with the hydrostatic pressure, especially in seemingly well studied semiconductors, are very essential in that case.

In the given work, the results of the quantitative analysis based on the experimental data on electronic transport under hydrostatic pressure are presented for the purpose of more deep study of the impurity energy spectrum in the arsenides, listed above. The next organization of the paper is obvious from the sector and subsector headers.

**1. Results and discussions**

The quantitative analysis of experimental results at room temperatures has been carried out for the two reasons. Firstly, in this case one may neglect a broadening of deep levels [11]. Besides, as it has been noted in [12-14], the influence of chaotic potential of defects is being enlarged with a lowering of the temperature because of free carrier depopulation. The last empties the chaotic traps and brings down the screening effect. All mentioned can cause errors in calculations of vacancy level positions in the energy scale relatively to conductivity band edge.

The known data on the dispersion law, effective mass of electrons at the bottom of a conductivity band $m_n$, width of the band-gap $\varepsilon_g$, pressure coefficient $d\varepsilon_g/dP$ have been used. Calculations have been done taking into account the statement about fixed level of a deep impurity energy, measured from electron affinity level [4-7], when hydrostatic pressure is being changed and with the use of next expressions:

$$n_{dj} = \frac{N_d}{1 + \beta \exp(\varepsilon_{dj}^* - \eta_j)} \qquad (1)$$

$$\ln\left(\frac{N_{dj}}{n_{dj}}-1\right)+\eta_j = \varepsilon_{d0}^* +\left(d\varepsilon_g/dP\right)^* \cdot P_j + \ln\beta \tag{1a}$$

$$N=n_j+n_{dj}=N_d+N_{sh}-N_a, \tag{2}$$

$$\frac{n_1-n_3}{n_1-n_2}=\frac{n_{d3}-n_{d1}}{n_{d2}-n_{d1}} \tag{2a}$$

$$\beta \exp\varepsilon_{d1} = \frac{1-A}{A\exp\left[(P_1-P_3)(d\varepsilon_g/dP)^*-\eta_3\right]-\exp\left[(P_1-P_2)(d\varepsilon_g/dP)^*-\eta_2\right]}, \tag{3}$$

where $$A = \frac{n_1-n_3}{n_1-n_2} \cdot \frac{\exp\left[(P_1-P_2)(d\varepsilon_g/dP)^*+\eta_1-\eta_2\right]-1}{\exp\left[(P_1-P_3)(d\varepsilon_g/dP)^*+\eta_1-\eta_3\right]-1}.$$

Indexes' 0', '1', '2', '3' correlate ambient and operative pressure values $P_1<P_2<P_3$ respectively, $\varepsilon_{dj}$ (j=1,2,3) - reduced energy between the bottom of the conductivity band and deep donor level of energy, $\eta_j$ (j=1,2,3) - reduced Fermi's energy, and $(d\varepsilon_g/dP)$ * - band-gap pressure coefficient at ambient pressure; $n_1$, $n_2$, $n_3$ and $n_{d1}$, $n_{d2}$, $n_{d3}$ - concentrations of electrons in the conductivity band and electrons, bound with the deep donors, $N_d$ - concentration of the deep donors, $\beta$ - parameter of spin degeneration.

In $CdSnAs_2$, InAs and $CdGeAs_2$, the spin-orbit splitting $\Delta$ and the band gap $\varepsilon_g$ are comparable each other. In this case it is necessary to start with three-band Kane model [15]. We used Fermi's two-parametrical integrals by the means the introduction of effective two-band parameters instead of Fermi's three-parametric integrals, which are not tabulated. The next parity [15] is applied:

$$m_n = \frac{\varepsilon_g \cdot (\varepsilon_g+\Delta)}{\varepsilon_g+2/3\cdot\Delta} \cdot \frac{\hbar^2}{2P_M^2} = \frac{3\hbar^2\varepsilon_g^*}{4P_M^2}, \tag{4}$$

where $P_M$ is a matrix element representing interaction between conduction and valence bands. The effective mass of electron and effective energy gap $\varepsilon_g^*$ are almost the same for above mentioned semiconductors:

$$\text{CdSnAs}_2: \varepsilon_g^* = (0.2 + 8.88 \cdot 10^{-16} P/P_0) \text{ eV}; \quad m_n/m_0 = 0.016 + 7.3 \cdot 10^{-17} P/P_0 \quad (5)$$

InAs: $\varepsilon_g^* = (0.322 + 7.6 \cdot 10^{-16} P/P_0)$ eV; $m_n/m_0 = 0.022 + 5.17 \cdot 10^{-17} P/P_0$ (6) CdGeAs$_2$: $\varepsilon_g^* = (0.405 + 6.41 \cdot 10^{-16} P/P_0)$ eV; $m_n/m_0 = 0.02 + 3.19 \cdot 10^{-17} P/P_0$ (7)

### 2.1. *n*-InAs

Typical dependences for Hall factor $R_H(P)$, resistancy $\rho(P)$ and mobility $\mu_H = |R_H|/\rho$ for single crystal *n*-InAs with concentration of excess donors of $\sim 10^{16}$ sm$^{-3}$ on pressure at 300 K are presented in figure 1 [16]. $R_H$ practically does not depend on pressure in the range till to $(2 \div 3)$GPa and then $|R_H|$ and $\rho$ increase with pressure by the exponential law at pressures up to $P \approx 6.5$ GPa (the beginning of polymorphic transition). Such a character of the dependence $R_H(P)$ in the range of 2.5÷6GPa is caused by the presence of the deep resonant donor impurity center - vacancies of arsenic [3, 17]. We notice, that in InAs, extremums of *L*-valley and *X*-valley are located above the $\Gamma$– valley extremum, by ~1 eV and ~1.5 eV, accordingly [18].

Our results concerning $R_H(P)$ (figure 1) for the sample with $n_0 = 1.84 \cdot 10^{16}$ sm$^{-3}$ are presented in figure 2. According to relations (1) - (3):

$$\ln\left(\frac{N_{rd}}{n_{rd}} - 1\right) + \eta = 13.43 - 4.26 \cdot P = \varepsilon_{rd0}^* + (d\varepsilon_g/dP)^* \cdot P + \ln\beta \quad (8)$$

Thus, we have obtained: $N_{rd} = 1.9 \cdot 10^{16}$ sm$^{-3}$, $N_{sh} - N_a = -0.06 \cdot 10^{16}$ sm$^{-3}$.

For $\beta = 1$, $\varepsilon_{dr} = (0.35 - 0.11 \cdot P)$ eV. Free electrons are completely localized at the donor centers and Fermi's energy is close to $\varepsilon_{rd}$ if $P > 4.5$GPa. Position of Fermi's level at $P > 5$GPa is stabilized with regard to resonant donor level (figure 2), and $\varepsilon_F > \varepsilon_{dr}$: $\varepsilon_F - \varepsilon_{dr} \approx 0.09$ eV.

### 2.2. *n*-CdSnAs$_2$ and *n*-CdGeAs$_2$

These materials, being cognate by their properties, belong to the most studied semiconductor group II-IV-V$_2$ [19, 20]. Their band structures are similar each another. However, electron mobility in *n*-CdGeAs$_2$ is much less in comparison with that in *n*-CdSnAs$_2$; also there exists some vagueness in experimental data about pressure dependence of resistivity in *n*-CdGeAs$_2$. Based on optical absorption, Faraday effect, photoconductivity and kinetic phenomena data, the conclusion about existence of additional subbands in the conductivity band of *n*-CdGeAs$_2$ and CdSnAs$_2$ has been drawn (see [19, 20] and references therein). Nevertheless, theoretical calculations [21, 22] have not confirmed any presence of additional minima in the conductivity band close to the main minimum, and observed features in experimental data are caused apparently by an impurity level.

Dependences $R_H$ (*P*) and $\rho$ (*P*) for *n*-InAs (figure 1) and *n*-CdSnAs$_2$ (figure 3) [23] are similar, that is due to the presence of a deep donor center which energy level falls in the continuum of the conductivity band in both materials; and a distinction between pressures, when these dependences are sharply growing, is caused by the fact that initial concentration of electrons in the investigated samples differ by two order of magnitude. In *n*-InAs - $n=1.84 \cdot 10^{16}$ sm$^{-3}$, *n*-CdSnAs$_2$ - $n=1.9 \cdot 10^{18}$ sm$^{-3}$ and their Fermi's energy values are $\varepsilon_F = -0.04$ eV and $\varepsilon_F = 0.19$ eV at atmospheric pressure, accordingly. Also, an increase in resistivity with the pressure in *n*-InAs is much more noticeable than that in *n*-CdSnAs$_2$ (figures 1 and 3).

In the investigated crystals *n*-CdSnAs$_2$, concentration of electrons practically does not depend on pressure untill *P*=1 GPa (a growth within the limits (1÷2) % is caused by volume-concentration effect [24]), and growth $\rho$ (*P*) is connected with a decrease of mobility of electrons [15].

Taking into account experimental dependence on the pressure for kinetic coefficients in the range of 1÷3.4 GPa (figure 3), initial concentration of electrons $n=1.9 \cdot 10^{18}$ sm$^{-3}$, dependence $\chi$ (*P*) [20] and calculated dependences $m_n(P)$ and $m_n(\eta)$, the dependence $n$ (*P*) and $\eta$ (*P*) has been defined with the use of known values band parameters and effective width of the forbidden band (6) in two-band approximation of Kane model [15], .

By means of dependences $n$ (*P*), $\eta$ (*P*), according to (1) - (3) one obtains:

$$\ln\left(\frac{N_{dr}}{n_{dr}} - 1\right) + \eta = 11.8 - 4.14 P = \varepsilon_{dr0}^* + (d\varepsilon_g/dP)^* P \qquad (9)$$

Thus, it is found: $N_{dr}=8\cdot10^{17}$ sm$^{-3}$, $N_{sh}-N_a=1.1\cdot10^{18}$ sm$^{-3}$. For $\beta=1$

$$\varepsilon_{dr}=(0.3 - 1.08\cdot10^{-15}\ P/P_0)\ \text{eV}, \qquad (10)$$

In figure 4, the results of the quantitative analysis are presented. Unlike the $n$-InAs sample with $N=1.84\cdot10^{16}$ sm$^{-3}$ and $N_{sh}-N_a = -0.06\cdot10^{16}$ sm$^{-3}$ (figure 1, 2), in the $n$-CdSnAs$_2$ sample with $N=1.9\cdot10^{18}$ sm$^{-3}$ we have $(N_{sh}-N_a)=1.1\cdot10^{18}$ sm$^{-3}$. In $n$-CdSnAs$_2$ $n \to N_{sh}-N_a$, $n_{dr} \to N_{dr}$, and in $n$-InAs $n \to 0$, $n_{dr} \to N$ if $P \to \infty$. The described situation explains, why $\rho$ in $n$-InAs increases more than by four order of magnitude prior to the beginning of phase transition, and $\rho$ in n-CdSnAs$_2$ increases only by ~5 times (figure 1, 3). Experimental data for $\rho(P)$ in $n$-CdGeAs$_2$ are inconsistent (figure 5) [25,26] up to $P=1$ GPa at 300 T. In these crystals we have $\rho(P)/\rho(0)$ =1.22÷1.24 ($P=1$ a GPa) [25] and $n = (10^{17}÷10^{18})$ sm$^{-3}$. An increase of $\rho$ at $0 \leq P \leq 1$ GPa, exactly like to that in n-CdSnAs$_2$, is caused by a decrease of mobility of electrons and matches with the pressure dependences of the band gap $\partial \varepsilon_g / \partial P=0.093$ eV·GPa$^{-1}$ [27] and static dielectric permeability. Thus, $\rho(P)$ for $n$-CdSnAs$_2$ and $\rho(P)$ for $n$-CdGeAs$_2$, doped with tellurium and indium, are close each other (figures 3 and 5). One can notice, that a decrease of $\chi$ with the pressure growth has an essential influence on $\rho(P) \sim \mu(P)^{-1}$ and should be taken into account for correct interpretation of experiment data.

In undoped $n$-CdGeAs$_2$ we have $\rho/\rho_0=7$ at $P=0.8$GPa [26] (figure 5). Such a strong increase of $\rho(P)$ in this material is well correlated with the presence of a deep donor level (vacancy of arsenic) near to conductivity band edge: $\varepsilon_d = (-0.05-0.093\cdot P)$eV. This result is in agreement with [3], where CdGeAs$_2$ exposed to irradiation has been investigated, and the presence of donors with $\varepsilon_d$ = -0.05 eV has been noted. For concentration of electrons at ambient pressure $n=10^{17}$ sm$^{-3}$ we have:

$$N_d = 1.65 \cdot 10^{18} \text{ sm}^{-3}, N_{sh}-N_a = -3 \cdot 10^{16} \text{ sm}^{-3}, \rho(P)/\rho_0 = 1+7.4 \cdot 10^{-14} \qquad (11)$$

The results of the quantitative analysis are presented in figure 6.

## 2.3. Gallium arsenide *n*-GaAs

Below, the results of the quantitative analysis on experimental data for pressure dependence of Hall coefficient $R_H$ and of resistancy in *n*-GaAs are presented at hydrostatic pressure from atmospheric to $P = 18$ GPa. It is known [28, 29], that в *n*-GaAs resistivity sharply changes with the growth of pressure twice: once near $P = 2$ GPa and secondly near $P = (5 \div 6)$ GPa, with the next saturation (figure 7). The Hall coefficient dependence on pressure is weak up to 2GPa, then the curve passes through an extremum at $P = (5 \div 6)$GPa and then becomes close to its atmospheric value. Such features of dependences $\rho(T)$ and $R_H(P)$ are caused by inter-valley $\Gamma$-$X$ transitions in the conductivity band (figure 7) that induce a leakage of electrons from $\Gamma$ - valley to $X$ - valley. The pressure factor of the band width between the edge of $X$ valley and the top of valent band is negative: $\Delta = d\varepsilon_x/dP = -14$ meV/GPa. The $\Gamma$-valley edge $\varepsilon_g$ is above $\varepsilon_X$ more than on 300 meV at 6 GPa ($d\varepsilon_\Gamma/dP = 94$ meV/GPa, $\varepsilon_X - \varepsilon_\Gamma = 360$ meV at $P=0$), and concentration of electrons in the $\Gamma$-valley $n_\Gamma \approx 0$ [28, 29]. Besides, from data on temperature dependences for $\rho(T)$ and $R_H(T)$ at ambient pressure in bulk *n* - GaAs with concentration of excess donors $N_d = 1.8 \cdot 10^{16} \div 5.5 \cdot 10^{17}$ sm$^{-3}$ [28], the energy level of the impurity centers is found to be $\varepsilon_{d1} = (0.15 - 1.1 \cdot 10^{-7} N_d^{1/3})$eV. Indeed, in the presence of deep donor which level of energy is located near to $\Gamma$-valley edge $\varepsilon_\Gamma - \varepsilon_{d1} \approx 150$ meV (when$P=0$), the calculated dependences $\rho(P)$ and $R_H(P)$ for $P < 6$ GPa fit with experimental data, but contradict to dependence $\rho(P)$ for $P>10$ GPa (figure 7). Concentration of electrons in $\Gamma$- valley at 10 GPa is $n_\Gamma \approx 0$ and a decrease of $\rho$ in the range of 10 GPa $<P<$ 18 GPa [29] also is observed (figure 7). It testifies the presence of deep donor energy level $\varepsilon_{d2}$ under the bottom of the $X$-valley $\varepsilon_X$, as the distance ($\varepsilon_X - \varepsilon_{d2}$) diminishes with the pressure growth, and therefore concentration of electrons in *X*-valley grows in turn. By the results of the quantitative analysis of dependence $\rho(P)$ at 10 GPa $<P<$18 GPa, it is found out (figure 8), that level of energy for this deep donor is settled down near to *Γ*- valley edge at ambient pressure. Calculations have been done with certain variation of total concentration in both valleys of the conductivity band ($10^{15} \div 10^{18}$) sm$^{-3}$ with the concept of independence of deep impurity centers on pressure [4-7], and with the use of relations (1) - (3). From experimental dependence $\rho(P)$ in the range 10 GPa $\leq P \leq$ 18 GPa, and for total concentration of electrons $10^{18}$ sm$^{-3}$ in *Γ*- and *X*-valleys, we obtain (figure 8):

$$\varepsilon_X - \varepsilon_{d2} = (289 - 1.42 \cdot 10^{-15} P/P_0) \text{ meV}, \tag{12}$$

and $\varepsilon_{d2} - \varepsilon_\Gamma = 70$ meV at *P*=0.

Thus, the deep donor center $\varepsilon_{d2}$, found out in *n*-GaAs at 10 GPa $\leq P \leq$ 18 GPa, is the second upper lying and partially populated "alternative" level of the double donor – the vacancy of arsenic. Concentration of compensating acceptors $N_a$ in the considered case submits the inequalilty $N_d < N_a < 2N_d$, that is caused by technological background of the sample material.

**Conclusion**

All specific conclusions have been made in the each chapter for every material correspondingly. Here we just summarize briefly the main points of the paper. Electron energy spectrum in undoped *n*-type bulk crystals of InAs, GaAs, CdSnAs$_2$ and CdGeAs$_2$ has been investigated on the basis of data on transport phenomena under hydrostatic pressure and their generalization and quantitative analysis. In GaAs for example, resistivity increases with increase of the pressure from the ambient value to *P*=2GPa as the consequence of inter-valley *Γ-X* transitions. At pressure 8GPa, the lowest in the conductivity band is the *Γ*-valley, which with the growth of *P* comes nearer to deep donor level and thereof a decrease of resistivity $\rho(P)$ is observed. In the interval 8 $<P<$18 GPa, $\rho(P)$ decreases by more than 5 times. From the comparison of the obtained results with data about influence of irradiation on impurity electronic spectrum of the set forth above semiconductors it is found that native defect - the vacancy of arsenic, corresponds to deep donor level. Positions of these levels concerning the edge of conductivity band and their pressure coefficient are obtained.

Figures and captures

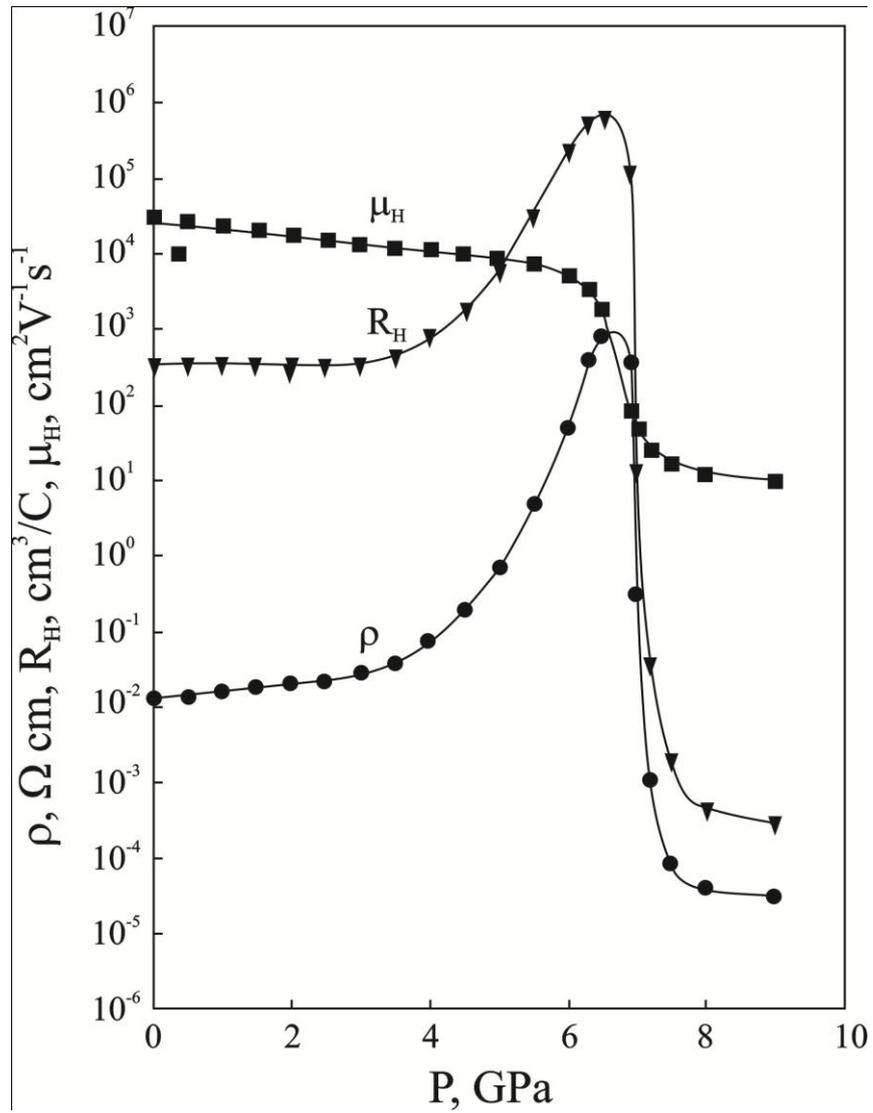

Figure 1. Dependences of resistivity, Hall coefficient and Hall mobility of electrons on pressure in the single crystal sample *n*-InAs at *T*=300 K with concentration of electrons at ambient pressure *n*=1.84·10$^{16}$ sm$^{-3}$.

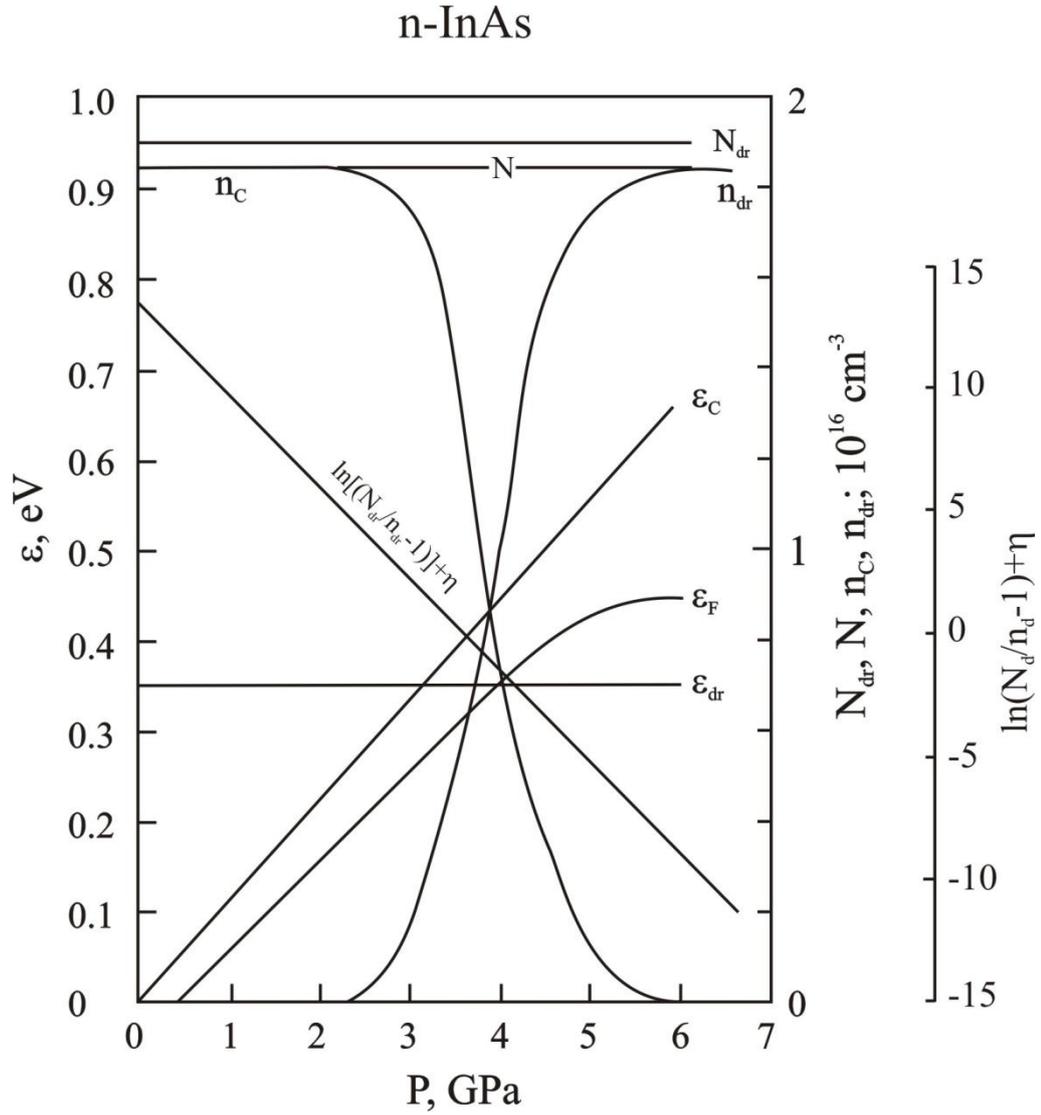

Figure 2. Dependences for energy of the bottom of conductivity band $\varepsilon_c$, the level of deep resonant donor $\varepsilon_{dr}$, Fermi's energy $\varepsilon_F$ relatively to $\varepsilon_{c0}$ ($P=0$), concentration of electrons in conductivity band $n$ and electrons located at the deep donor centers $n_{dr}$, and function $\ln[(N_{dr}/n_{dr}-1)]+\eta$ on the pressure in $n$-InAs at $T=300$ K with concentration of electrons at ambient pressure $n=1.84\cdot10^{16}$ sm$^{-3}$. $N=n+n_{dr}$, $N_{dr}$ - concentration of the deep donor centers.

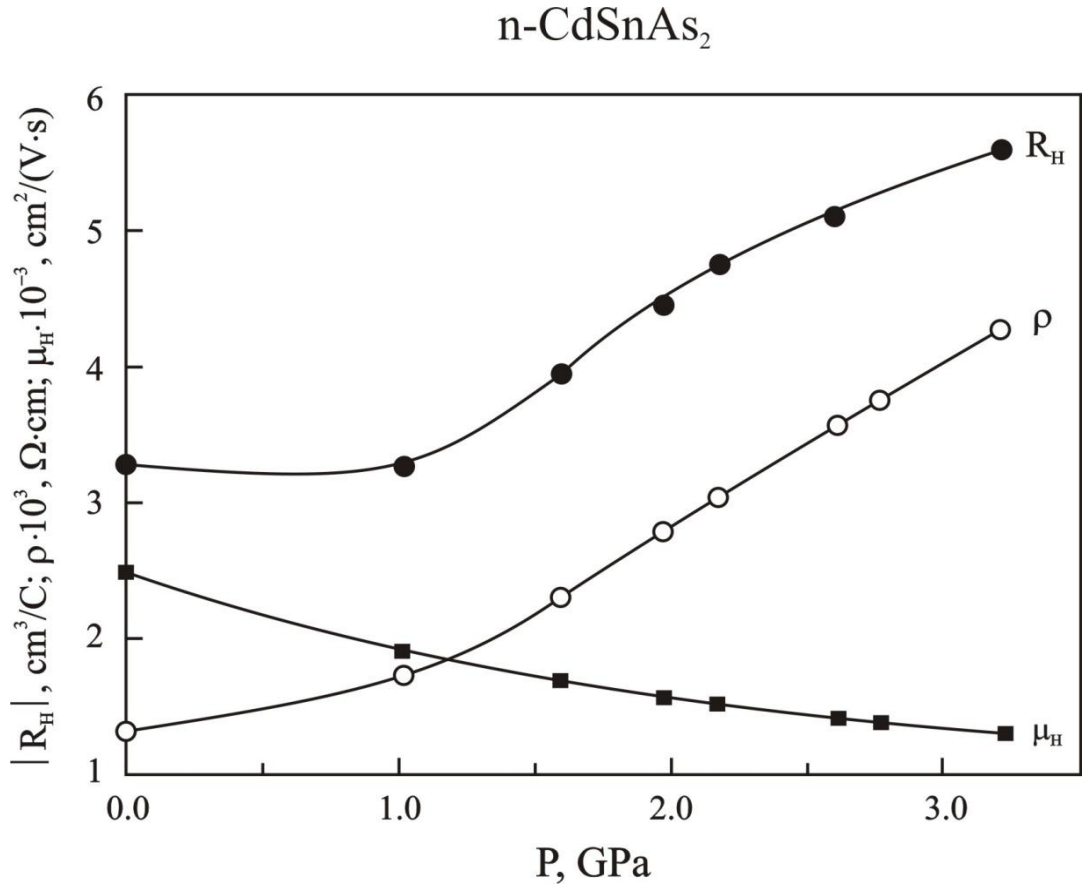

Figure 3. Dependences of resistivity, Hall coefficient and Hall mobility of electrons on pressure in the single crystal *n*-CdSnAs$_2$ at *T*=300 K with concentration of electrons at ambient pressure $n=1.9 \cdot 10^{18}$ sm$^{-3}$.

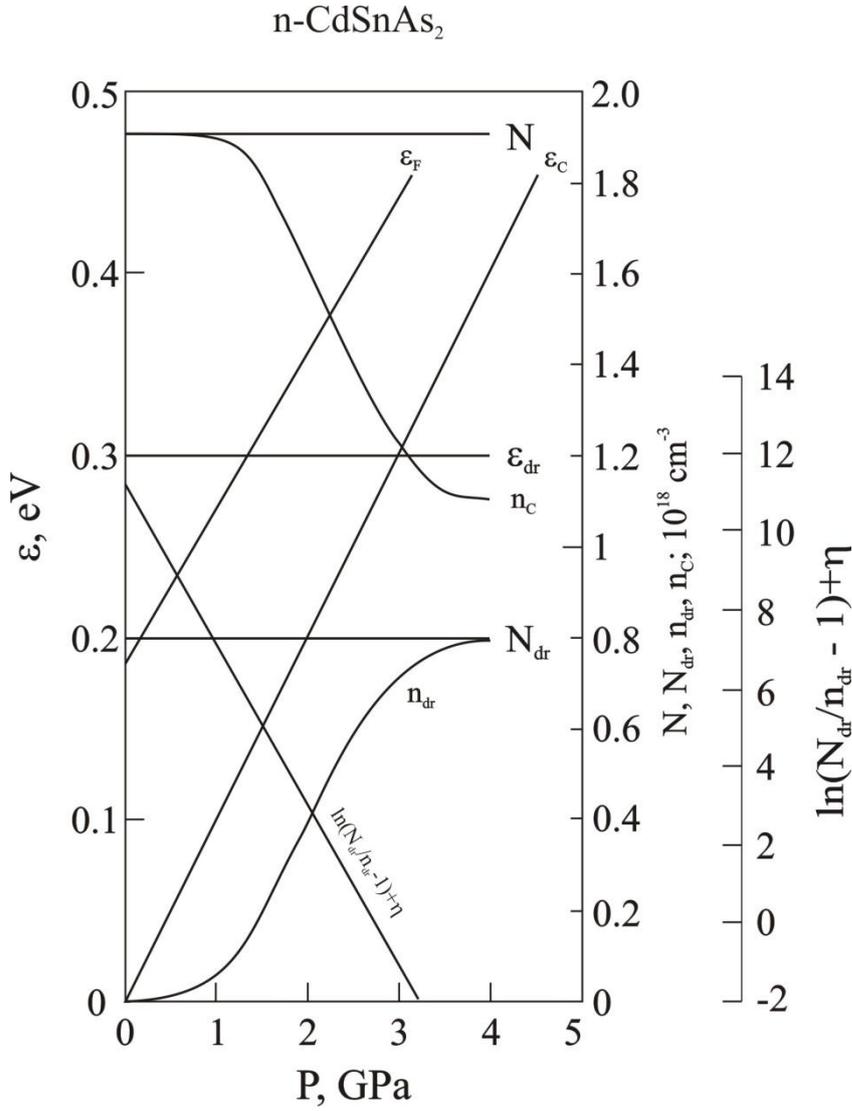

Figure 4. Pressure dependences for bottom of the conductivity band $\varepsilon_C$, Fermi's energy $\varepsilon_F$, energy of deep resonant donor $\varepsilon_{dr}$ relatively to $\varepsilon_{C0}$ ($P=0$), and concentrations of electrons in conductivity band $n$, and electrons located at deep donor centers $n_{dr}$, and function $\ln[(N_{dr}/n_{dr}-1)]+\eta$ for $n$-CdSnAs$_2$ with concentration of electrons at atmospheric pressure $n=1.9 \cdot 10^{18}$ sm$^{-3}$ and $T=300$ K. $N_{dr}$ and $N_{sh}$ – concentrations of deep and shallow donor, $N_a$ – concentration of compensating acceptors ($N=n+n_{dr}=N_{dr}+N_{sh}-N_a$).

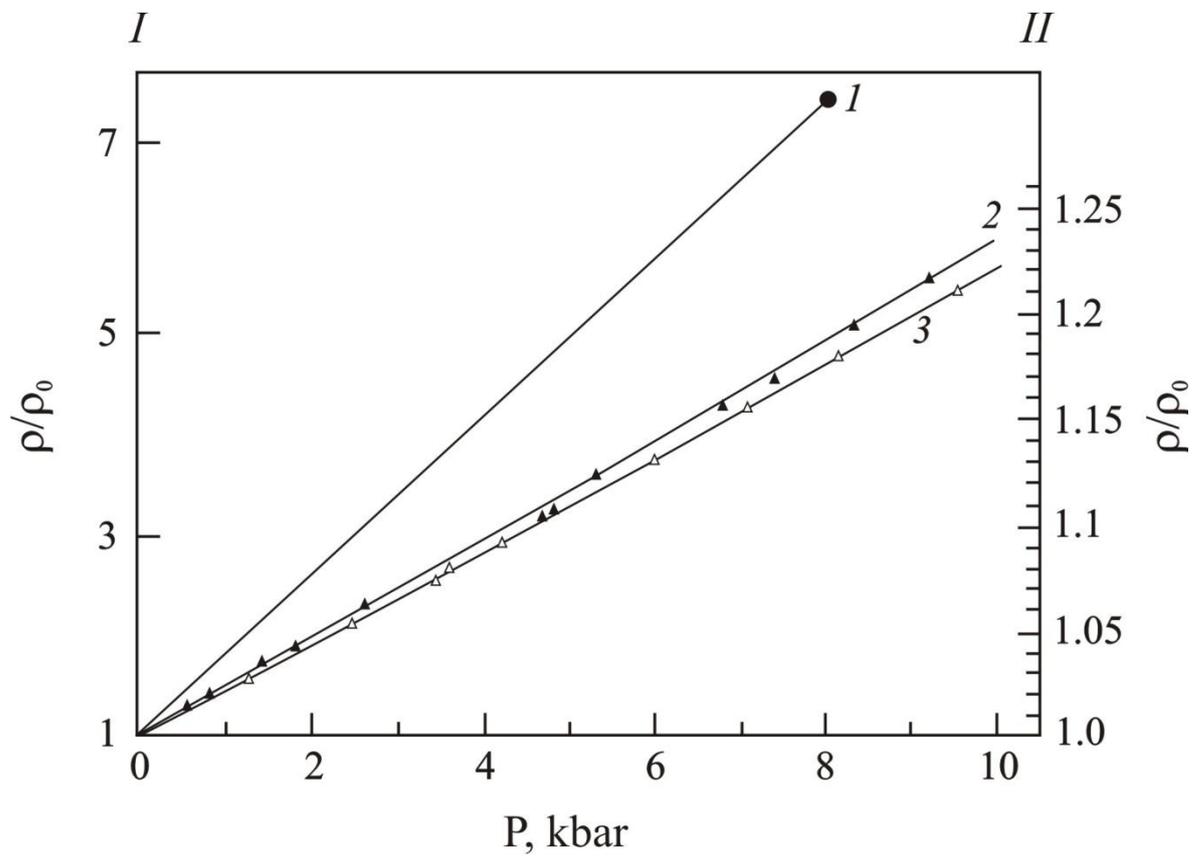

Figure 5. Dependences of resistivity on pressure in undoped (1) (axis I) [26] and doped by tellurium (2) and indium (3) (axis II) [25] $n$-CdGeAs$_2$ at 300 K.

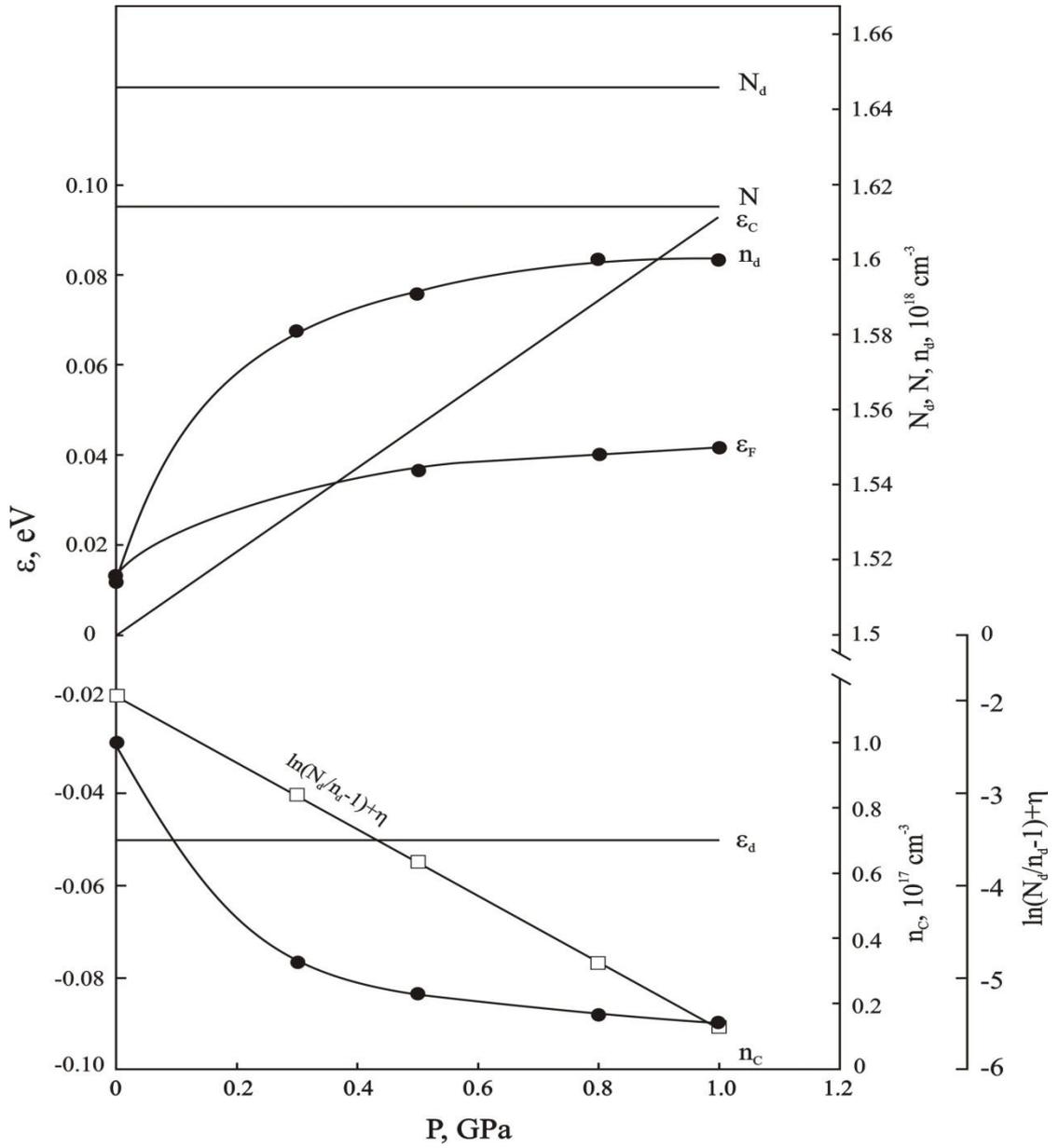

Figure 6. Pressure dependences for bottom of the conductivity band $\varepsilon_C$, Fermi's energy $\varepsilon_F$, energy level of the deep resonant donor centers $\varepsilon_d$, relatively to $\varepsilon_{C0}$ ($P=0$) and concentration of electrons in the conductivity band $n$, electrons on the deep donor centers $n_d$, and function $\ln[(N_d/n_d-1)]+\eta$ for $n$-CdGeAs$_2$ with concentration of electrons at atmospheric pressure $n=10^{17}$ sm$^{-3}$ at $T=300$ K. $N_{dr}$ and $N_{sh}$ – concentrations of deep and shallow donors, $N_a$ – concentration of compensating acceptors, ($N=n+n_d=N_d+N_{sh}-N_a$).

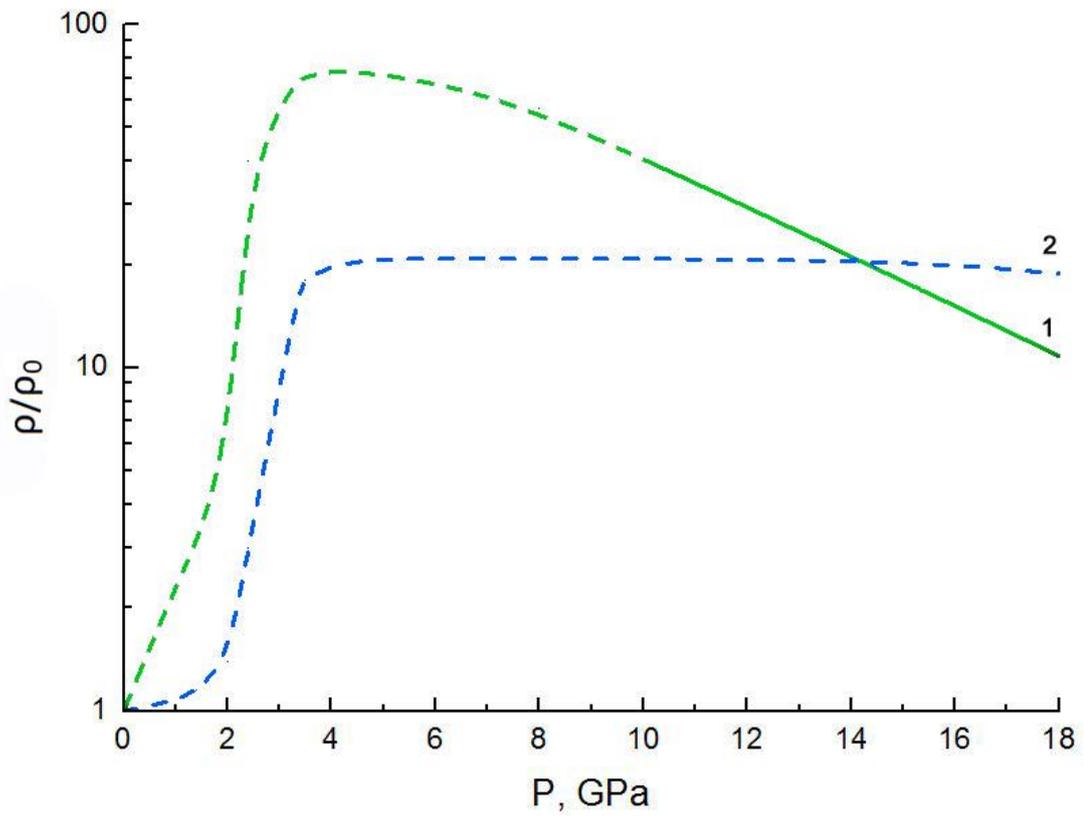

Figure 7. Dependences of the normalised resistivity $\rho / \rho_o$ on pressure for *n*-GaAs: a solid line - experiment [29], dashed lines - calculation for two values of energy of the deep double donor.

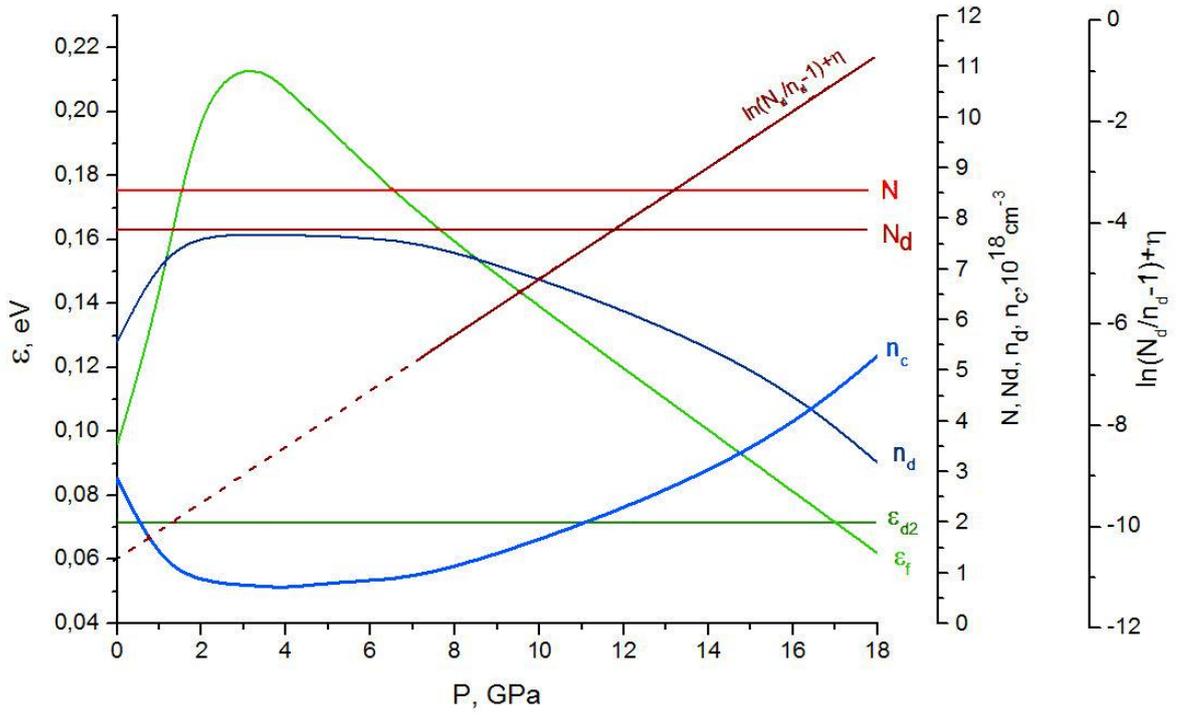

Figure 8. The calculated pressure dependences for *n*-GaAs for Fermi energy $\varepsilon_F$, the energy of the deep donor $\varepsilon_{d2}$ relatively to the edge of $\Gamma$-valleys at ambient pressure, concentration of electrons located at deep donors $n_d$ and their total concentration in $\Gamma$- and $X$ -valleys $n= n_X+n_\Gamma$, and function $\ln[(N_d/n_d-1)]+\eta$; $N = n_\Gamma + n_X+n_d$.